\documentclass{pasj01}
\Received{$\langle$February 9, 2012$\rangle$}
\Accepted{$\langle$acception date$\rangle$}
\Published{$\langle$publication date$\rangle$}

\begin{document}

\title{%
 Iron Fluorescent Line Emission from the mCvs and Hard X-ray Emitting Symbiotic Stars as a Source of the  Iron Fluorescent Line Emission from the Galactic Ridge
}

\author{%
Romanus~\textsc{Eze},\altaffilmark{1,2}
Kei~\textsc{Saitou},\altaffilmark{1,3}
and 
Ken~\textsc{Ebisawa}\altaffilmark{1,3}
}
\altaffiltext{1}{%
Japan Aerospace Exploration Agency, Institute of Space and Astronautical Science,\\
3-1-1 Yoshinodai, Chuo-ku, Sagamihara, Kanagawa 252-5210}
\altaffiltext{2}{%
Department of Physics and Astronomy, University of Nigeria, Nsukka, Nigeria}
\altaffiltext{3}{%
Department of Astronomy, Graduate School of Science, The University of Tokyo,
7-3-1 Hongo, Bunkyo-ku, Tokyo 113-0033}
\email{eze@ac.jaxa.jp or romanus.eze@unn.edu.ng}


\maketitle

\begin{abstract}
The Galactic Ridge X-ray Emission (GRXE) spectrum has strong iron emission lines at 6.4, 6.7, 
and 7.0~keV, each corresponding to the neutral (or low-ionized), He-like, and H-like iron ions. 
The 6.4~keV fluorescence line is due to irradiation of  neutral (or low ionized) 
material (iron) by hard X-ray sources, indicating uniform presence of the cold matter
in the Galactic plane. In order to resolve origin of the cold fluorescent matter,
we examined  the contribution of the 6.4~keV line emission from white dwarf surfaces 
in the hard X-ray emitting symbiotic stars (hSSs) and magnetic cataclysmic variables (mCVs)
to the GRXE. 
In our spectral analysis of 4~hSSs and 19~mCVs observed with Suzaku, we were able to resolve 
the three iron emission lines. 
We found that the equivalent-widths (EWs) of the 6.4~keV lines of hSSs are systematically higher 
than those of mCVs, such that the average EWs of hSSs and mCVs are $179_{-11}^{+46}$~eV and 
$93_{-3}^{+20}$~eV, respectively. 
The EW of hSSs compares favorably with the typical EWs of the 6.4~keV line in the GRXE of 90--300~eV 
depending on Galactic positions.
Average 6.4~keV line luminosities of the hSSs and mCVs are 
$9.2\times 10^{39}$ and $1.6\times 10^{39}$~photons~s$^{-1}$, respectively, indicating that 
hSSs are intrinsically more efficient 6.4~keV line emitters than mCVs.
We compare expected contribution of the 6.4 keV lines from mCVs with the observed GRXE 6.4 keV line flux  in the direction of $(l,b) \approx (28.5\arcdeg, 0\arcdeg$).  
We conclude that almost all the 6.4 keV line flux in GRXE may be explained by mCVs within  current
undertainties of the stellar number densities, while contribution from hSSs may not be negligible.
\end{abstract}

\section{Introduction}
Presence of the seemingly extended hard X-ray emission from the Galactic Ridge has been 
known since early 1980's (Galactic Ridge X-ray Emission; GRXE: 
\cite{worrall1982,warwick1985,koyama1986}).
Strong iron K-line emission at $\sim$6.7~keV in the GRXE indicates its thermal plasma origin 
(e.g., \cite{koyama1986,yamauchi1993}).
More precise iron line diagnostics of the GRXE has been made possible with X-ray CCD cameras 
on-board ASCA \citep{kaneda1997} and Chandra \citep{ebisawa2005}. 
These instruments revealed that the line centroid energies in the GRXE are systematically 
lower than 6.7~keV (the energy expected from He-like ion in thermal equilibrium plasma), 
which imply that either the line emission is from non-ionization equilibrium plasma or there 
is an additional 6.4~keV fluorescent line emission. 
Suzaku, for the first time, resolved the GRXE iron line emission into three narrow lines, ones 
from neutral or low ionized (6.4~keV), He-like (6.7~keV), and H-like (7.0~keV) ions 
\citep{ebisawa2008}, concluding that the GRXE iron line emission is both from hot thermal plasmas 
and fluorescence by cold materials.

Regarding the origin of the GRXE, there is a strong argument in favor of collection of faint 
point sources as opposed to the diffused emission (e.g., 
\cite{revnivtsev2006,krivonos2007,revnivtsev2009,revnivtsev2010} and references therein), although the 
question remains ``what are these Galactic point sources?'' Candidate point sources for 
the GRXE are cataclysmic variables (CVs) and active binaries (ABs). 
CVs are known to have such strong emission lines and hard spectra, but most of them are 
brighter than $\sim$$10^{31}$~erg~s$^{-1}$, and their population may not be sufficient to 
account for all the GRXE. 
\citet{revnivtsev2009} proposed that ABs dimmer than $\sim$$10^{31}$~erg~s$^{-1}$ are likely 
candidates to account for the majority of the GRXE. 
However, ABs are well known to have thermal but much softer continuum spectra than CVs. 
\citet{yuasa2010b} proposed that intermediate polars (IPs), which are a subclass of magnetic CVs 
(mCVs), are main sources of the hard X-ray emission of the GRXE, giving support to the point source 
scenario for the origin of the GRXE.
\citet{yuasa2010b} concluded that combination of IPs and ABs will explain most of the 6.7~keV 
and 7.0~keV emission lines in the GRXE as well as the continuum emission above 20~keV, whereas 
an \textit{ad hoc} additional 6.4~keV line component is needed to explain the entire GRXE by 
the point source model. 

This X-ray fluorescence is, on the other hand, believed by some authors to be due to irradiation 
of the molecular clouds by X-ray photons or it may be as result of cosmic-ray particle bombardment 
\citep{koyama1986,dogiel1998,murakami2000,valinia2000,koyama2007,yusefzadeh2007,dogiel2009,capelli2011}. 
This is also likely, since Galactic $\gamma$-ray \textit{diffuse} emission above $\sim$100~keV 
is successfully explained by the cosmic-rays and interstellar matter interaction model 
(e.g., \cite{strong2005}). 

In this paper, we study origin of the 6.4~keV emission line in the GRXE, examining if 
this emission could be fully resolved by collection of point sources. 
We focused on hard X-ray emitting symbiotic stars (hSSs) and magnetic CVs (mCVs) observed with 
the Suzaku satellite \citep{mitsuda2007}, since they are known to be significant 6.4~keV line 
emitters (e.g., \cite{ezuka1999,yuasa2010a,luna2007,smith2008,kennea2009,eze2011}). 
We studied 4~hSSs and 19~mCVs (one polar and 18~IPs), all observed with the Suzaku satellite, 
and estimated their contributions to the 6.4~keV line emission flux of the GRXE. 
Our goal is to determine if they are the main sources of the GRXE 6.4~keV line emission flux, 
or if some additional sources are required.

\section{Data Selection}
Our target sources, hSSs and  mCVs, were selected based on the fact that they have been 
observed with Suzaku and have strong Fe~K$\alpha$ emission lines with hard X-ray emission above 20~keV. 
%
All the four hard X-ray emitting symbiotic stars, SS73~17, RT~Cru, T~CrB, and CH~Cyg observed with Suzaku were selected. 
In selecting the mCVs (polars and IPs), we used a CV catalog \citep{ritter2008} and the IP 
catalog\footnote{%
  http://asd.gsfc.nasa.gov/Koji.Mukai/iphome/catalog/alpha.html
}. 
Five sources in the catalog, AE~Aqr, AM~Her, GK~Per, 1RXS~J070407.9+26250, and 1RXS~J180340.0+40121 
were dropped, even though observed with Suzaku, because they appear to had been too faint during 
their observations or have particular emission mechanism (AE~Aqr: e.g., \cite{wynn1997}). 
A total of 23~sources were thus selected (table~\ref{t1}).

\section{Data Analysis and Results}
Analysis of our data were done using version~2 of the standard Suzaku pipeline products, and 
the HEASoft\footnote{%
  See http://heasarc.gsfc.nasa.gov/lheasoft/ for details. 
} version~6.10. 
In majority of the sources we used $250\arcsec$~radius to extract all events for the XIS 
detector for the production of the source spectra but in some cases where the $250\arcsec$~radius 
overlaps with the calibration sources at the corners, we adjusted the radius accordingly. 
The XIS background spectra were extracted with $250\arcsec$~radius with no apparent sources 
and were offset from both the source and corner calibrations. 
The radius was also adjusted accordingly in some cases where it overlaps with the calibration 
sources at the corners. 
Response Matrices File and Ancillary Response File were generated for the XIS detector using the 
FTOOLS \texttt{xisrmfgen} and \texttt{xissimarfgen} \citep{ishisaki2007}, respectively. 
Suzaku XIS~0, 2, and 3 have front-illuminated (FI) chips with similar features, so we merged the 
spectra of XIS~0 and 3, which we hereafter refer to as XIS~FI (XIS~2 has been out of service 
since November 9, 2006 due to an anomaly). 
XIS~1 is back-illuminated (BI), and we hereafter refer it to as XIS~BI. 

In the HXD PIN detector analysis, we used the non-X-ray background files and response matrix 
files appropriate for each observation provided by the Suzaku team. 
We used the \texttt{mgtime} FTOOLs to merge the good time intervals to get a common value for 
the PIN background and source event files. 
The source and background spectra extraction were done for each observation using the 
\texttt{xselect} filter time file routine. 
We corrected for the dead time of the observed spectra using the \texttt{hxddtcor} in the 
Suzaku FTOOLS. 
According to the standard analysis procedure, exposure time for all observations for the derived 
background spectra were increased by a factor of 10 to take care of the event rate in the PIN 
background event file which is made 10 times higher than the real background for suppression of 
the Poisson errors. 
We assumed a cosmic background model obtained with the HEAO-1 satellite \citep{boldt1987}. 

Spectral analysis of all observations were performed using XSPEC version~12.7. 
We modeled the spectrum using absorbed bremsstralung model with three Gaussian lines for 
the three Fe~K$\alpha$ emission lines to measure the iron line fluxes. 
We assumed two types of absorption by full-covering and partial covering matter. 
Since we were primarily interested in the ion lines, our fitting covers 3--10~keV for the XIS~BI, 
3--12~keV for the XIS~FI and 15--40~keV for the HXD~PIN. 
We ignored energy range below 3~keV in the XIS~FI and BI detector to avoid intrinsic absorption 
which is known to affect data at this energy range, and energies above 10~keV were ignored for 
XIS~BI because the instrument background is higher compared to the XIS~FI detectors. 
We also ignored energy range above 40~keV in the HXD~PIN detector in order to obtain high signal-to-noise 
ratio signals. 

The three Fe lines, neutral or low ionized (6.4~keV), He-like (6.7~keV), and H-like (7.0~keV) ions, 
were clearly resolved in all the sources except IGR~J17303--0601 where we were unable to detected 
the H-like (7.0~keV) significantly but the other two lines were detected. 
Spectra of all the sources can be found in an
earlier work by \citet{eze2015} and 
spectral parameters were shown in table~\ref{t2}. 

Furthermore, in order to determine average spectra of the hSSs and the mCVs, 
we used \texttt{addascaspec} to average the spectra of hSSs and mCVs (as well as responses).
We used the same model to produce the spectra for the average hSSs and average mCVs. 
The spectra and spectral parameters for the average hSSs and average mCVs were presented in 
figure~1 and table~\ref{t2}, respectively. 

We detected strong 6.4~keV iron line emission in the average hSSs spectrum with an equivalent 
width (EW) of $179_{-11}^{+46}$~eV and in the average mCVs spectrum with $93_{-3}^{+20}$~eV. 
We have found that the 6.4~keV line EW is much stronger in hSSs than in mCVs, which 
suggests that hSSs can be strong candidates of the GRXE 6.4~keV line emission. 
For comparison, 6.4~keV iron line EWs of the GRXE are of 90--390~eV, depending on the Galactic 
locations \citep{yamauchi2009}.

\section{Discussion}

\subsection{The 6.4~keV Line Emission}
The 6.4 keV fluorescence line emission is usually due to irradiation
of the neutral (or low ionized) material (iron) by a hard
X-ray source. \citet{eze2015} used these sources used
in this work to study the origin of the 6.4 keV fluorescence line
and found that the emission could be partly from the reflection
of hard X-rays from the accretion disks/white dwarf surfaces and
from the absorption column.
Generally, for all compact objects that accretes matter such as white dwarfs, neutron stars, and 
black holes, there are often emission of the Fe~K$\alpha$ fluorescence line as well as Compton 
reflection which signifies  presence of the surrounding cold matter. 

hSSs and mCVs are both binary systems in which white dwarfs accrete matter from their companions. 
The CVs are semi-detached systems in which the secondary star fills its Roche lobe and starts 
transferring mass into the lobe of the compact white dwarf primary. 
The transferred material has too much angular momentum to fall directly onto the surface of 
the white dwarf, but instead builds an accretion disk, which spirals round the white dwarf 
(e.g., \cite{warner1995}). 
On the other hand, SSs are interacting binaries formed from a red giant star and a hot degenerate 
companion which accretes mass from the stellar wind of the red giant. 
Such a nebulae often formed surrounding the system is typically detected via various optical 
emission lines, whereas accretion disks may not be formed \citep{kenyon1986}.
Therefore, at least for hSSs, the 6.4~keV fluorescent emission lines are considered to be mainly 
from white dwarf surfaces.
Hard X-rays emitted in the vicinity of the white dwarfs irradiate the white dwarf surfaces (e.g., 
\cite{luna2007,eze2010}), leading to the emission of the Fe~K$\alpha$ fluorescence line. 

We found that the EWs of the 6.4~keV line of hSSs are systematically higher than those of mCVs, 
such that the average EW of hSSs and mCVs are $179_{-11}^{+46}$~eV and $93_{-3}^{+20}$~eV, 
respectively (table~\ref{t2}). 
In order to see if hSSs are truly more efficient 6.4~keV line emitters than mCVs, we estimated
6.4~keV photon luminosities for those hSSs and mCVs whose distances are known (table~\ref{t3}). 
As a result, we found that the 6.4~keV line luminosities of hSSs are systematically higher than 
those of mCVs; that of the average hSSs is $9.2\times 10^{39}$~photons~s$^{-1}$, and that of 
the average mCVs is $1.6\times 10^{39}$~photons~s$^{-1}$. 
We note that one of the hSSs, RT~Cru, has a high luminosity of $30\times 10^{39}$~photons~s$^{-1}$, 
however, if we isolate this source, the average luminosity of hSSs will become 
$2.2\times 10^{39}$~photons~s$^{-1}$, which will still be higher than that of the mCVs.

We suppose there is a reason why the 6.4~keV line luminosities of hSSs are significantly higher 
than those of mCVs, even though hSSs may lack accretion disks.
Many hSSs are believed to have a cocoon of gas coming from the red giant companion and surrounding 
the white dwarf (e.g., \cite{luna2007,eze2010,eze2011}).
Hence, there is a possibility that the  additional 6.4~keV line emission is from irradiation of 
the thick cold absorbing circumstellar gas partially covering the hard X-ray source. 
In fact, our spectral analysis indicate that the average hSSs spectrum has significantly higher 
hydrogen circumstallar partial covering fraction of $0.70\pm0.01$, compared to that 
of the average mCVs, $0.44\pm0.01$ with similar hydrogen column density of partial covering 
matter (table~\ref{t2}).

\subsection{The Contribution of the 6.4~keV Line Emission of hSSs and mCVs to that of the GRXE}
If the Galactic 6.4~keV line source has the line luminosity ($L_{\mathrm{6.4}}$~photons~s$^{-1}$)
and number the density ($n$~cm$^{-3}$), the 6.4~keV line emissivity may be expressed as 
\begin{eqnarray}
  \frac{nL_{\mathrm{6.4}}}{4\pi} \ \textrm{photons~s$^{-1}$~cm$^{-3}$~str$^{-1}$}
  \label{eq1}
\end{eqnarray}
and the observed 6.4~keV line flux in photons~s$^{-1}$~cm$^{-2}$~str$^{-1}$ will be
\begin{eqnarray}
  F_{\mathrm{6.4}}= \int \frac{n\left( x\right) L_{\mathrm{6.4}}}{4\pi} dx
    \approx  \frac{{\langle L_{\mathrm{6.4}}\rangle}}{4\pi} \int  n(x) dx, 
  \label{eq2}
\end{eqnarray}
where the integration is made along the line of sight, and $\langle L_{\mathrm{6.4}}\rangle$ is the average
6.4 keV line luninosity either from hSSs or mCVs
 (see also \cite{yamauchi2009}).
In order to be precise, we will have to take into account luminosity functions  of the hSSs and mCV to estimate
 $\langle L_{\mathrm{6.4}}\rangle$.  However, it is known that contribution of CVs to the GRXE is
limited to a narrow range of the luminosities (e.g., \citet{sazonov2006,warwick2014}).  Thus, we approximate 
 $\langle L_{\mathrm{6.4}}\rangle$ with the values estimated in the previous section
from the current Suzaku samples,  $\langle L_{\mathrm{6.4}}\rangle  \sim9.2 \times 10^{39}$ photons s$^{-1}$ for hSSs and $\sim1.6 \times 10^{39}$ photons s$^{-1}$ for mCVs.

Let's consider the measurement of the 6.4~keV photon flux at the position 
$(l,b) \approx (28.5\arcdeg, 0\arcdeg)$, 
$(8\pm 2)\times 10^{-5}$~photons~s$^{-1}$~cm$^{-2}$~deg$^{-2} \approx 0.26$~photons~s$^{-1}$~cm$^{-2}$~str$^{-1}$ 
\citep{ebisawa2008}.  We are going to estimate the contributions from mCVs using  (\ref{eq2}).

Following  \citet{sazonov2006,revnivtsev2007} and \citet{warwick2014}, we take the model of the stellar mass distribution on the Galactic plane,
\begin{equation}
\rho = \rho_{0, disk} \;  \exp \left[ -\left(\frac{R_m}{R}\right)^3 -\frac{R}{R_{disk}} \right],\label{massdensity}
\end{equation}
where $R_m$ and $R_{disk}$ represents the inner cut-off radius and e-fold disk radius, respectively.
These authors claim that almost all the total GRXE flux is explained assuming  the stellar density in the form of 
(\ref{massdensity}).  However, 
we found that the three authors adopt slightly different disk size and normalization, while
they use the same outer disk radius, $R_{max}$ = 10 kpc:
\citet{sazonov2006}  takes $\rho_{0, disk}=0.61 M_\odot$ pc$^{-3}$, $R= 3$ kpc, $R_{disk}=3$ kpc.
\citet{revnivtsev2007}  takes $\rho_{0, disk}=5.5 M_\odot$ pc$^{-3}$, $R_m= 2.5$ kpc, $R_{disk}~2.2$ kpc.
\citet{warwick2014}  takes the same $R_m$ and $R_{disk}$ as \citet{revnivtsev2007}, but $\rho_{0, disk}=1.6 M_\odot$ pc$^{-3}$.
Because of these differences, the total surface mass density integrated in the direction of $l=28.5\arcdeg$  is
significantly different: 1200 \citep{sazonov2006}, 1900  \citep{warwick2014} or 6700 \citep{revnivtsev2007}  $M_{\odot}$ pc$^{-2}$ (see Fig. 2). This may not be very surprising, since only less than  half of the GRXE is resolved into point sources
in this direction \citep{ebisawa2005}, and  estimate of the number of remaining extremely dim stars has a large uncertainty.
If we assume the expected spatial number density of CVs normalized by stellar mass, $1.2 \times 10^{-5} M_\odot^{-1}$ \citep{sazonov2006} we may calculate the expected 6.4 keV line photon flux from CVs; 0.19, 0.31 or 1.1 photons s$^{-1}$ cm$^{-2}$ str$^{-1}$,
depending on the three different stellar mass density estimates. Obviously, the latter two are over-estimates, exceeding the
observed value.  In any case, we conclude that the observed 6.4 keV line photon flux in the direction of $l=28.5\arcdeg$  
may be explained totally by CVs, within the current uncertainty of the stellar densities.

What about the contribution from hSSs?
We have very few number of known symbiotic stars 200
\citep{belczynski200}, of which only five are hSSs (see \citet{kennea2009}). This makes it difficult to estimate the space density of the hSSs and we therefore require more discoveries of hSSs in the Galaxy in other to make a proper estimate of the space density.
However, we believe that there are many more hSSs that are yet to
be discovered, hence hSSs could contribute a significant percentage
of the GRXE 6.4 keV line flux.

In summary, we  conclude that the GRXE 6.4~keV line flux is  primarily explained by  mCVs.
Contribution from 
the hSSs may not be neglected, since hSSs are intrinsically strong 6.4~keV line emitters.
Taking account of contributions from other types of white dwarf X-ray binaries, the GRXE 6.4~keV 
line emission may be fully explained as being from a large number of accreting white dwarfs 
in the Galactic plane, most of which has yet to be identified.

In order to confirm our model, further work should be done to search for more, still dimmer 
6.4~keV line emitting sources in the Galactic plane, which may be hSSs, mCVs, or other types 
of white dwarf binaries.
We hope that next generation Galactic surveys by the incoming hard X-ray satellite missions 
such as Spectrum-Roentgen-Gamma" (SRG) satellite will detect a large number of such sources that would account for most of the GRXE 6.4~keV line emission.

\bigskip 

We acknowledge the Suzaku team for providing data and some relevant files used in the analysis 
of this work. 
R.\,E. is very grateful to the Japan Society for the Promotion of Science (JSPS) for financial 
support under the JSPS Invitation Fellowship (Long Term), the Nigerian TETFund for National Research grant support and ISAS/JAXA, Sagamihara Campus for 
hosting him. 
This research made use of data obtained from Data ARchives and Transmission
System (DARTS), provided by Center for Science-satellite Operation and Data
Archives (C-SODA) at ISAS/JAXA. Some part of this work were\footnote{%
 Reprinted from Publication of New Astronomy, Vol. 36(2015), Author:  R.N.C., Eze, Title: On the origin of the iron fluorescent line emission from the Galactic Ridge/ Page No. 64 - 69, Copyright (2015), with permission from Elsevier}


%

%
\begin{figure*}[h]
\begin{center}
\begin{minipage}{70truemm}
  \FigureFile(80mm,80mm){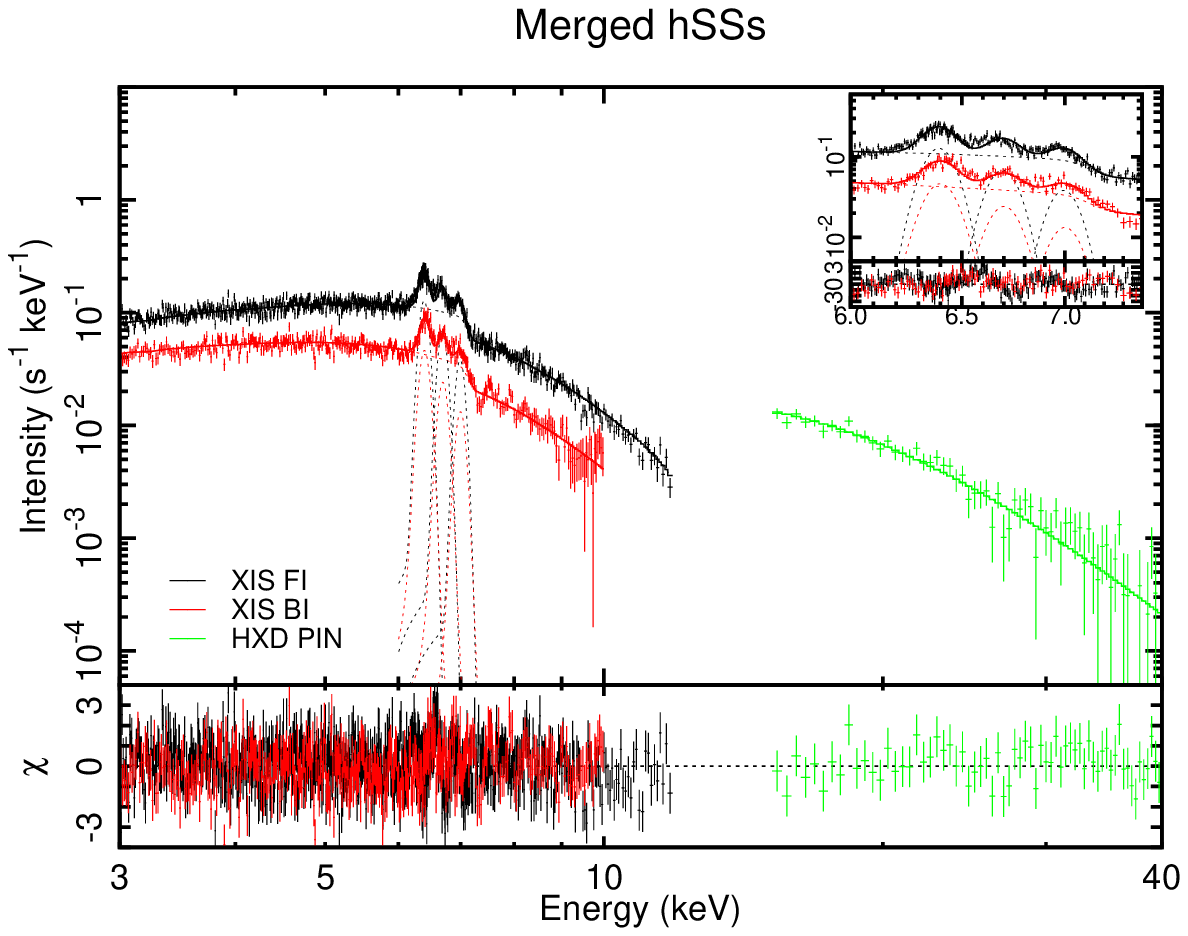}
\end{minipage}
\hspace{20truemm}
\begin{minipage}{70truemm}
  \FigureFile(80mm,80mm){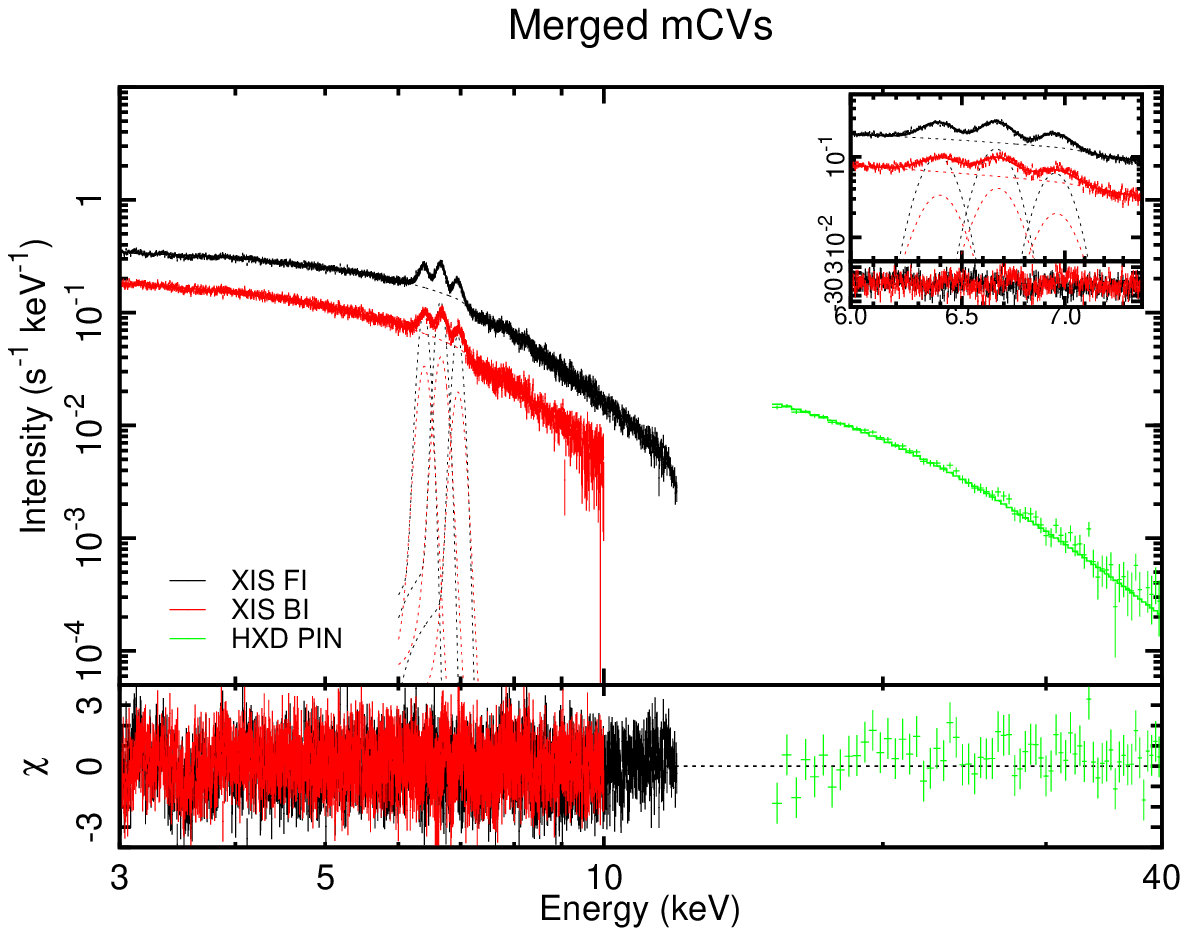}
\end{minipage}
\end{center}
  \caption{Spectra of the average symbiotic stars and the average magnetic cataclysmic variables.  
           In the upper panel, the data and the best-fit model are shown by crosses and solid 
           lines, respectively. 
           Each spectral component is represented by dotted lines. 
           In the lower panel, the ratio of the data to the best-fit model is shown by crosses. 
           The inset in the upper panel is an enlarged view for the Fe~K$\alpha$ complex lines.}
~\ref{f1}
\end{figure*}
\begin{figure*}[h]
\begin{center}
\begin{minipage}{70truemm}
  \FigureFile(80mm,80mm,angle=270){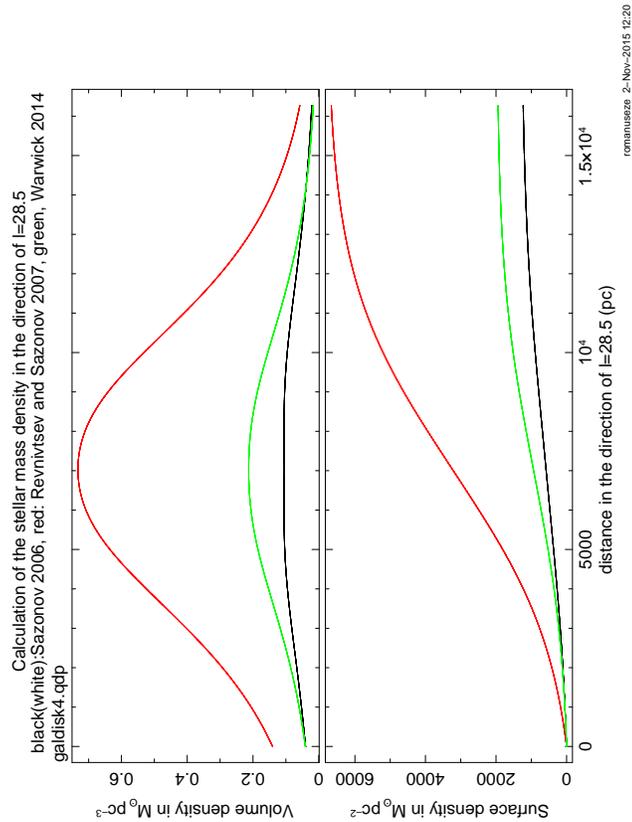}
\end{minipage}
\end{center}
  \caption{Calculation of the stellar mass density}
  \end{figure*}

\clearpage

%
\begin{table}[h]
\begin{center}
  \caption{The symbiotic stars, polars, and intermediate polars used in this work}
  \label{t1}
\begin{tabular}{lccc}
  \hline
  Source Name & ObsID & Obs. start (UT) & Exp. \\ 
   & & Date / Time & (ks) \\
  \hline
  \multicolumn{4}{c}{Symbiotic Stars} \\ 
  \hline
  CH~Cyg & 
    400016020 & 2006-05-28 / 07:28 & \phantom{0}35.2 \\
  T~CrB & 
    401043010 & 2006-09-06 / 22:44 & \phantom{0}46.3 \\
  RT~Cru & 
    402040010 & 2007-07-02 / 12:38 & \phantom{0}50.9 \\
  SS73~17 & 
    403043010 & 2008-11-05 / 16:30 & \phantom{0}24.9 \\
  \hline
  \multicolumn{4}{c}{Polars} \\ 
  \hline
  V1432~Aql & 
    403027010 & 2008-04-16 / 21:33 & \phantom{0}24.9 \\
  \hline 
  \multicolumn{4}{c}{Intermediate Polars} \\ 
  \hline 
  NY~Lup & 
    401037010 & 2007-02-01 / 15:17 & \phantom{0}86.8 \\
  RX~J2133.7$+$5107 & 
    401038010 & 2006-04-29 / 06:50 & \phantom{0}62.8 \\
  EX~Hya & 
    402001010 & 2007-07-18 / 21:23 & \phantom{0}91.0 \\
  V1223~Sgr & 
    402002010 & 2007-04-13 / 11:31 & \phantom{0}46.2 \\
  MU~Cam & 
    403004010 & 2008-04-14 / 00:55 & \phantom{0}50.1 \\
  V2400~Oph & 
    403021010 & 2009-02-27 / 11:42 & 110.0 \\
  YY~Dra & 
    403022010 & 2008-06-15 / 18:37 & \phantom{0}27.4 \\
  TV~Col & 
    403023010 & 2008-04-17 / 18:00 & \phantom{0}30.1 \\
  V709~Cas & 
    403025010 & 2008-06-20 / 10:24 & \phantom{0}33.3 \\
  IGR~J17303$-$0601 & 
    403026010 & 2009-02-16 / 10:09 & \phantom{0}27.7 \\
  IGR~J17195$-$4100 & 
    403028010 & 2009-02-18 / 11:03 & \phantom{0}26.9 \\ 
  BG~CMi & 
    404029010 & 2009-04-11 / 12:11 & \phantom{0}45.0 \\
  PQ~Gem & 
    404030010 & 2009-04-12 / 13:46 & \phantom{0}43.2 \\ 
  TX~Col & 
    404031010 & 2009-05-12 / 19:19 & \phantom{0}51.1 \\
  FO~Aqr & 
    404032010 & 2009-06-05 / 08:14 & \phantom{0}33.4 \\
  AO~Psc & 
    404033010 & 2009-06-22 / 11:50 & \phantom{0}35.6 \\
  IGR~J00234$+$6141 & 
    405022010 & 2010-06-25 / 00:06 & \phantom{0}77.4 \\
  XY~Ari & 
    500015010 & 2006-02-03 / 23:02 & \phantom{0}93.6 \\
  \hline
\end{tabular}
\end{center}
\end{table}

\clearpage
\begin{table*}[h]
  \vspace{20truemm}
\rotatebox{90}{\begin{minipage}{240mm}
\begin{center}
  \caption{The Fe lines fit parameters.\footnotemark[${*}$]}
  \label{t2}
\begin{scriptsize}
\begin{tabular}{lcccccccccccccccccc}
 \hline
  Source Name &
    $N_{\mathrm{H}}^{\mathrm{f}}$ & $N_{\mathrm{H}}^{\mathrm{p}}$ & $C$ &
    $kT$ & $F_{\mathrm{cont}}$ &
    $E_{\mathrm{6.4}}$ & $F_{\mathrm{6.4}}$ & EW$_{\mathrm{6.4}}$ &
    $E_{\mathrm{6.7}}$ & $F_{\mathrm{6.7}}$ & EW$_{\mathrm{6.7}}$ &
    $E_{\mathrm{7.0}}$ & $F_{\mathrm{7.0}}$ & EW$_{\mathrm{7.0}}$ \\
  \hline
  \multicolumn{15}{c}{Symbiotic Stars} \\
  \hline 
  CH~Cyg & 
    $22.0\pm7.0$ & $99_{-37}^{+26}$ & $0.91\pm0.04$ & 
    $6_{-2}^{+3}$ & $9.2\pm0.5$ & 
    $6.41\pm0.02$ & $20.2\pm0.6$ & $580_{-65}^{+424}$ & 
    $6.59\pm0.05$ & $5.9\pm1.8$ & $111_{-35}^{+95}$& 
    $6.81_{-0.07}^{+0.12}$ & $1.8\pm0.7$ & \phantom{0}$66_{-38}^{+87}$ \\
  T~CrB & 
    $17.7\pm2.2$ & $37\pm6$ & $0.71\pm0.06$ & 
    $19_{-2}^{+3}$ & $14.7\pm1.3$ & 
    $6.43\pm0.01$ & $9.6\pm0.6$ & $117_{-46}^{+48}$ & 
    $6.72\pm0.02$ & $6.5\pm0.5$ & $85_{-12}^{+40}$ & 
    $7.01\pm0.02$ & $5.3\pm0.3$ & \phantom{0}$104_{-41}^{+38}$ \\
  RT~Cru & 
    $3.3\pm0.4$ & $58\pm9$ & $0.46\pm0.05$ & 
    $29_{-5}^{+9}$ & $12.0\pm1.1$ & 
    $6.38\pm0.01$ & $11.0\pm0.9$ & $174_{-30}^{+38}$ & 
    $6.64\pm0.01$ & $5.8\pm0.5$ & $52_{-15}^{+34}$ & 
    $6.96\pm0.01$ & $3.2\pm0.3$ & \phantom{0}$51_{-25}^{+17}$ \\
  SS73~17 & 
    $9.2\pm0.5$ & $34\pm5$ & $0.65\pm0.05$ & 
    $38\pm6$ & $4.9\pm1.0$ & 
    $6.38\pm0.01$ & $7.6\pm0.6$ & $185_{-63}^{+120}$ & 
    $6.67\pm0.01$ & $7.7\pm0.2$ & $158_{-41}^{+136}$ & 
    $6.95\pm0.01$ & $4.4\pm0.1$ & \phantom{0}$93_{-40}^{+116}$ \\
  \hline 
  \multicolumn{15}{c}{Polars} \\
  \hline 
  V1432~Aql & 
    $3.0\pm0.5$ & $61\pm9$ & $0.56\pm0.05$ & 
    $15\pm3$ & $14.7_{-2.9}^{+3.8}$ & 
    $6.42\pm0.01$ & $6.5\pm0.7$ & $91_{-20}^{+33}$ & 
    $6.67\pm0.02$ & $5.5\pm0.5$ & $84_{-8}^{+40}$ & 
    $7.00\pm0.03$ & $2.0\pm0.4$ & \phantom{0}$42_{-34}^{+29}$ \\
  \hline 
  \multicolumn{15}{c}{Intermediate Polars} \\
  \hline 
  NY~Lup & 
    $0.6\pm0.5$ & $32\pm8$ &  $0.35\pm0.03$ & 
    $40\pm7$ & $9.6_{-0.3}^{+0.5}$ & 
    $6.40\pm0.01$ & $7.1\pm0.3$ & $118_{-12}^{+34}$ & 
    $6.66\pm0.01$ & $5.9\pm0.3$ & $94_{-9}^{+43}$ & 
    $6.94\pm0.01$ & $4.7\pm0.2$ & \phantom{0}$70_{-17}^{+28}$ \\
  RX~J2133.7+5107 & 
    $2.3\pm0.3$ & $73\pm10$ & $0.46\pm0.05$ & 
    $29_{-6}^{+10}$ & $8.9\pm0.9$ & 
    $6.42\pm0.01$ & $6.3\pm0.5$ & $147_{-26}^{+46}$ & 
    $6.68\pm0.02$ & $3.1\pm0.3$ & $52_{-14}^{+32}$ & 
    $6.98\pm0.02$ & $2.3\pm0.3$ & \phantom{0}$57_{-26}^{+16}$ \\
  EX~Hya & 
    $0.7\pm0.5$  & $98\pm16$ &  $0.54\pm0.06$ & 
    $10\pm1$ & $20.8\pm0.1$ & 
    $6.41\pm0.01$ & $3.3\pm0.2$ & $28\pm3$ & 
    $6.66\pm0.02$ & $2.9\pm0.3$ & $32\pm1$ & 
    $6.95\pm0.01$ & $1.1\pm0.3$ & \phantom{0}$109_{-4}^{+5}$ \\
  V1223~Sgr & 
    $2.3\pm0.2$ & $72\pm5$ & $0.41\pm0.03$ & 
    $25\pm3$ & $34.5\pm0.3$ & 
    $6.38\pm0.01$ & $15.7\pm0.7$ & $89_{-8}^{+17}$ & 
    $6.67\pm0.01$ & $11.3\pm0.6$ & $59_{-8}^{+18}$ & 
    $6.95\pm0.01$ & $8.3\pm0.5$ & \phantom{0}$45_{-12}^{+14}$ \\
  MU~Cam & 
    $1.8\pm1.3$ & $35\pm14$ & $0.48\pm0.06$ & 
    $29_{-8}^{+14}$ & $3.6_{-0.4}^{+0.8}$ & 
    $6.41\pm0.01$ & $2.9\pm0.3$ & $160_{-32}^{+70}$ & 
    $6.68\pm0.02$ & $2.5\pm0.2$ & $104_{-22}^{+71}$ & 
    $6.98\pm0.01$ & $2.4\pm0.2$ & \phantom{0}$102_{-36}^{+58}$ \\
  V2400~Oph & 
    $1.4\pm0.2$ & $64\pm7$ & $0.36\pm0.04$ & 
    $18\pm2$ & $16.8\pm0.3$ & 
    $6.39\pm0.01$ & $9.7\pm0.5$ & $112_{-10}^{+26}$ & 
    $6.68\pm0.01$ & $7.1\pm0.3$ & $77_{-7}^{+24}$ & 
    $6.96\pm0.01$ & $4.7\pm0.3$ & \phantom{0}$57_{-13}^{+20}$ \\
  YY~Dra & 
    $1.1\pm0.1$ & $102_{-30}^{+28}$  & $0.52\pm0.08$ & 
    $17\pm1$ & $8.2\pm0.1$ & 
    $6.41\pm0.02$ & $2.2\pm0.3$ & $49_{-22}^{+34}$ & 
    $6.69\pm0.01$ & $6.6\pm0.4$ & $157_{-19}^{+52}$ & 
    $6.99\pm0.02$ & $4.4\pm0.4$ & \phantom{0}$173_{-92}^{+63}$ \\
  TV~Col & 
    $4.1\pm0.2$ & $36\pm7$ & $0.34\pm0.05$ & 
    $27\pm2$ & $13.6\pm1.2$ & 
    $6.41\pm0.01$ & $8.7\pm0.3$ & $114_{-12}^{+39}$ & 
    $6.69\pm0.02$ & $9.7\pm0.2$ & $122_{-11}^{+45}$ & 
    $6.98\pm0.01$ & $6.7\pm0.1$ & \phantom{0}$99_{-21}^{+41}$ \\
  V709~Cas & 
    $0.5\pm0.4$ & $51\pm18$ & $0.26\pm0.07$ & 
    $26_{-5}^{+7}$ & $11.2_{-1.3}^{+2.0}$ & 
    $6.41\pm0.02$ & $6.3\pm0.5$ & $108_{-18}^{+33}$ & 
    $6.69\pm0.02$ & $2.8\pm0.4$ & $56_{-15}^{+29}$ & 
    $6.99\pm0.02$ & $2.8\pm0.3$ & \phantom{0}$48_{-24}^{+27}$ \\
  IGR~J17303$-$0601 & 
    $1.9\pm0.7$ & $106_{-28}^{+25}$ & $0.56_{-0.12}^{+0.09}$ & 
    $29\pm7$ & $10.5\pm1.7$ & 
    $6.39\pm0.02$ & $4.6\pm0.7$ & $89_{-16}^{+20}$ & 
    $6.67\pm0.04$ & $2.3\pm0.5$ & $45_{-13}^{+18}$ & 
    --- & --- & --- \\
  IGR~J17195$-$4100 & 
    $1.4\pm0.1$ & $21_{-5}^{+8}$ & $0.27\pm0.02$ & 
    $26\pm5$ & $9.6_{-0.5}^{+0.2}$ & 
    $6.40\pm0.01$ & $5.9\pm0.4$ & $113_{-20}^{+40}$ & 
    $6.69\pm0.01$ & $4.9\pm0.4$ & $90_{-17}^{+39}$ & 
    $6.97\pm0.02$ & $4.0\pm0.4$ & \phantom{0}$82_{-26}^{+40}$ \\
  BG~CMi & 
    $5.4_{-1.0}^{+2.9}$ & $43\pm24$ & $0.39\pm0.09$ & 
    $19_{-4}^{+5}$ & $8.8_{-1.4}^{+2.7}$ & 
    $6.40\pm0.02$ & $3.6\pm0.4$ & $86_{-15}^{+17}$ & 
    $6.65\pm0.03$ & $2.4\pm0.3$ & $57_{-14}^{+23}$ & 
    $7.00\pm0.04$ & $1.3\pm0.3$ & \phantom{0}$39_{-14}^{+23}$ \\
  PQ~Gem & 
    $2.0\pm0.4$ & $64\pm13$ & $0.46\pm0.07$ & 
    $18\pm6$ & $7.7\pm0.2$ & 
    $6.39\pm0.01$ & $5.0\pm0.6$ & $137_{-39}^{+28}$ & 
    $6.66\pm0.03$ & $1.7\pm0.4$ & $32_{-25}^{+60}$ & 
    $6.93\pm0.04$ & $1.1\pm0.4$ & \phantom{0}$23_{-23}^{+37}$ \\
  TX~Col & 
    $1.8\pm0.6$ & $59\pm16$ & $0.50\pm0.08$ & 
    $12_{-3}^{+4}$ & $5.2_{-0.1}^{+0.3}$ & 
    $6.38\pm0.02$ & $1.6\pm0.3$ & $47_{-33}^{+51}$ & 
    $6.68\pm0.02$ & $2.4\pm0.3$ & $83_{-30}^{+71}$ & 
    $6.95\pm0.02$ & $1.6\pm0.2$ & \phantom{0}$49_{-36}^{+52}$ \\
  FO~Aqr & 
    $10.2\pm0.2$ & $143_{-25}^{+56}$  & $0.61\pm0.07$ & 
    $27\pm2$ & $37.0\pm1.0$ & 
    $6.37\pm0.01$ & $28.4\pm1.0$ & $149_{-30}^{+39}$ & 
    $6.65\pm0.01$ & $16.6\pm1.0$ & $80_{-54}^{+33}$ & 
    $6.92\pm0.02$ & $11.1\pm1.0$ &\phantom{0}$49_{-22}^{+26}$ \\
  AO~Psc & 
    $4.2\pm0.2$ & $40\pm5$ & $0.57\pm0.06$ & 
    $17\pm1$ & $13.8_{-0.1}^{+0.8}$ & 
    $6.37\pm0.01$ & $8.9\pm0.3$ & $91_{-16}^{+38}$ & 
    $6.66\pm0.02$ & $13.0\pm0.2$ & $136_{-16}^{+47}$ & 
    $6.94\pm0.01$ & $7.4\pm0.2$ & \phantom{0}$59_{-21}^{+24}$ \\
  IGR~J00234$+$6141 & 
    $0.4\pm0.5$ & $32\pm6$ & $0.30\pm0.04$ & 
    $24\pm7.2$ & $2.1_{-0.4}^{+1.6}$ & 
    $6.40\pm0.03$ & $1.2\pm0.2$ & $111_{-39}^{+87}$ & 
    $6.65\pm0.04$ & $6.2\pm0.1$ & $49_{-36}^{+102}$ & 
    $6.97\pm0.03$ & $0.6\pm0.1$ & \phantom{0}$60_{-43}^{+97}$ \\
  XY~Ari & 
    $6.9\pm0.2$ & $26\pm4$ & $0.49\pm0.05$ & 
    $28\pm4$ & $5.1\pm0.6$ & 
    $6.39\pm0.01$ & $2.5\pm0.7$ & $85_{-15}^{+32}$ & 
    $6.68\pm0.01$ & $3.1\pm0.3$ & $112_{-12}^{+33}$ & 
    $6.97\pm0.01$ & $1.7\pm0.4$ & \phantom{0}$72_{-17}^{+31}$ \\
  \hline 
  \multicolumn{15}{c}{Merged Spectra} \\ 
  \hline 
  Merged~hSSs &
    $5.6\pm0.2$ & $50\pm2$ & $0.70\pm0.01$ & 
    $20\pm1$ & $10.7\pm1.3$ & 
    $6.408_{-0.005}^{+0.002}$ & $7.7\pm0.7$ & $179_{-11}^{+46}$ & 
    $6.671\pm0.007$ & $5.2\pm0.3$ & $88_{-8}^{+43}$ & 
    $6.969_{-0.008}^{+0.013}$ & $3.8\pm0.3$ & \phantom{0}$74_{-17}^{+28}$ \\
  Merged~mCVs &
    $1.6\pm0.1$ & $66\pm1$ & $0.44\pm0.01$ & 
    $18\pm1$ & $12.3\pm0.4$ & 
    $6.403_{-0.001}^{+0.003}$ & $5.8\pm0.1$ & $93_{-3}^{+20}$ & 
    $6.671_{-0.002}^{+0.001}$ & $7.7\pm0.1$ & $114_{-4}^{+11}$ & 
    $6.957_{-0.002}^{+0.003}$ & $4.2\pm0.1$ & \phantom{0}$62_{-4}^{+7}$ \\
  \hline
  \multicolumn{15}{@{}l@{}}{\hbox to 0pt{\parbox{255mm}{\footnotesize 
  \noindent 
  \footnotemark[$*$] 
    Parameters are 
    the hydrogen column density of the full-covering and the partial-covering matter in units 
    of $10^{22}$~cm$^{-2}$ ($N_{\mathrm{H}}^{\mathrm{f}}$ and $N_{\mathrm{H}}^{\mathrm{p}}$), 
    the covering fraction of the partial-covering matter ($C$), 
    the continuum temperature in keV ($kT$), 
    the continuum flux in $10^{-3}$~photons~s$^{-1}$~cm$^{-2}$ ($F_{\mathrm{cont}}$), 
    the center energies of 6.4, 6.7, and 7.0~keV lines in keV ($E_{\mathrm{6.4}}$, 
    $E_{\mathrm{6.7}}$, and $E_{\mathrm{7.0}}$), 
    the line fluxes in $10^{-5}$~photons~s$^{-1}$~cm$^{-2}$ ($F_{\mathrm{6.4}}$, 
    $F_{\mathrm{6.7}}$, and $F_{\mathrm{7.0}}$), and 
    the equivalent widths in eV (EW$_{\mathrm{6.4}}$, EW$_{\mathrm{6.7}}$, and EW$_{\mathrm{7.0}}$). 
    EX~Hya and YY~Dra do not have both absorptions ($N_{\mathrm{H}}^{\mathrm{f}}$ and 
    $N_{\mathrm{H}}^{\mathrm{p}}$), 
    IGR~J17195$-$4100 has no $N_{\mathrm{H}}^{\mathrm{f}}$, 
    while TV~Col and FO~Aqr have no $N_{\mathrm{H}}^{\mathrm{p}}$. 
 }\hss}}
\end{tabular}
\end{scriptsize}
\end{center}
\end{minipage}}
\end{table*}

\clearpage

\begin{table}[h]
\begin{center}
  \caption{6.4~keV line luminosities of the sources.\footnotemark[${*}$]}
  \label{t3}
\begin{tabular}{lcccccccccccccccccc}
  \hline
  Source Name & Luminosity & Distance  & Ref. \\ 
   &  ($10^{39}$~photons~s$^{-1}$) & (pc) \\ 
  \hline
  \multicolumn{4}{c}{Symbiotic Stars} \\ 
  \hline
  CH~Cyg & 
    \phantom{0}1.52 & \phantom{0}250 & [1] \\
  T~CrB & 
    \phantom{0}2.90 & \phantom{0}500 & [2] \\
  RT~Cru & 
    30.0 & 1500 & [3] \\
  SS73~17 & 
    \phantom{0}2.29 & \phantom{0}500 & [4] \\
  \hline
  \multicolumn{4}{c}{Polars} \\ 
  \hline
  V1432~Aql & 
    \phantom{0}0.42 & \phantom{0}230 & [5] \\
  \hline 
  \multicolumn{4}{c}{Intermediate Polars} \\ 
  \hline 
  NY~Lup & 
    \phantom{0}6.05 & \phantom{0}840 & [5, 6] \\
  RX~J2133.7$+$5107 & 
    \phantom{0}--- & \phantom{0}--- & --- \\
  EX~Hya & 
    \phantom{0}0.02 & \phantom{00}65 & [7] \\
  V1223~Sgr & 
    \phantom{0}5.27 & \phantom{0}527 & [2, 7] \\
  MU~Cam & 
    \phantom{0}--- & \phantom{0}--- & --- \\
  V2400~Oph & 
    \phantom{0}2.93 & \phantom{0}500 & [8] \\
  YY~Dra & 
    \phantom{0}0.06 & \phantom{0}155 & [9] \\
  TV~Col & 
    \phantom{0}1.44 & \phantom{0}370 & [10] \\
  V709~Cas & 
    \phantom{0}0.48 & \phantom{0}250 & [6] \\
  IGR~J17303$-$0601 & 
    \phantom{0}--- & \phantom{0}--- & --- \\
  IGR~J17195$-$4100 & 
    \phantom{0}0.09 & \phantom{0}110 & [11] \\ 
  BG~CMi & 
    \phantom{0}1.10 & \phantom{0}500 & [9] \\
  PQ~Gem & 
    \phantom{0}0.97 & \phantom{0}400 & [9] \\ 
  TX~Col & 
    \phantom{0}0.48 & \phantom{0}500 & [9] \\
  FO~Aqr & 
    \phantom{0}5.50 & \phantom{0}400 & [8] \\
  AO~Psc & 
    \phantom{0}0.67 & \phantom{0}250 & [9] \\
  IGR~J00234$+$6141 & 
    \phantom{0}0.41 & \phantom{0}530 & [12] \\
  XY~Ari & 
    \phantom{0}0.22 & \phantom{0}270 & [13] \\
  \hline 
  \multicolumn{4}{c}{Average Spectra} \\ 
  \hline 
  hSSs~Average & 
    \phantom{0}9.20 & \phantom{0}--- & --- \\
  mCVs~Average\footnotemark[$\dagger$] & 
    \phantom{0}1.63 & \phantom{0}--- & --- \\
  \hline
  \multicolumn{4}{@{}l@{}}{\hbox to 0pt{\parbox{100mm}{\footnotesize 
  \noindent 
  \footnotemark[$*$] 
    References for the source distance: 
    [1]~\citet{sokokenyon2003}; [2]~\citet{kennea2009}; [3]~\citet{luna2007}; 
    [4]~\citet{smith2008}; [5]~\citet{dematino2006}; [6]~\citet{balow2006};
    [7]~\citet{beuerman2003}; [8]~\citet{suleimanov2005}; [9]~\citet{patterson1994}; 
    [10]~\citet{mcauthur2001}; [11]~\citet{masetti2006}; [12]~\citet{bonnet-bidaud2007}; 
    [13]~\citet{littlefair2001}. \\ 
  \footnotemark[$\dagger$] 
    There are no estimated distances for RX~J2133.7$+$5107, MU~Cam, and IGR~J17303$-$0601, 
    hence they were not considered in the estimation of the average luminosity of the mCVs.
 }\hss}}
\end{tabular}
\end{center}
\end{table}

\end{document}